\documentclass[12pt]{article}
\usepackage{latexsym}

\renewcommand{\thefootnote}{\fnsymbol{footnote}}

\begin{document}
\setlength{\baselineskip}{5mm}

\begin{titlepage}
\setcounter{page}{0}
\begin{flushright}
EPHOU 00-004\\
May 2000
\end{flushright}
\vspace*{20mm}
\begin{center}
{\Large A Conserved Energy Integral for Perturbation Equations \\
 \ \\
        in the Kerr-de Sitter Geometry}

\vspace{10mm}
{\bf Hiroshi Umetsu\footnote{
e-mail address: umetsu@particle.sci.hokudai.ac.jp}}\\
\vspace{5mm}
{\em Department of Physics, Hokkaido University \\
Sapporo, 060-0810 Japan}\\
\end{center}
\vspace{10mm}

\begin{abstract}
The analytic proof of mode stability of the Kerr black hole was provided 
by Whiting. In his proof, the construction of a conserved quantity 
for unstable mode was crucial.
We extend the method of the analysis for the Kerr-de Sitter geometry.
The perturbation equations of massless fields in the Kerr-de Sitter 
geometry can be transformed into Heun's equations which have four 
regular singularities.
In this paper we investigate differential and integral transformations 
of solutions of the equations.
Using those we construct a conserved quantity for unstable modes 
in the Kerr-de Sitter geometry, and discuss its property.
\end{abstract}
\end{titlepage}

\renewcommand{\thefootnote}{\arabic{footnote}}
\setcounter{footnote}{0}

\newpage

\section{Introduction}
\hspace*{\parindent}
\setcounter{equation}{0}
One of the most non-trivial aspects of perturbation equations for 
the Kerr-de Sitter geometry is the separability of the radial and 
angular parts.
Carter first showed that the perturbation equation for a scalar field 
is separable in the Kerr-Newman-de Sitter geometry~\cite{Carter}.
This observation was extended for spin $1/2$, electromagnetic fields, 
gravitational perturbations and gravitinos for the Kerr geometries and 
even for the Kerr-de Sitter class of geometries.
These perturbation equations are called Teukolsky equations~\cite{Teukolsky}.
Except for electromagnetic and gravitational perturbations, 
the separability persist even for the Kerr-Newman-de Sitter solutions.

An important application of the separability is the proof of the stability 
of the black hole.
The proof of mode stability of the Kerr black hole was provided by 
Whiting~\cite{Whiting} in 1989.
The proof is more complicated than one for the Schwarzschild black hole, 
because in the Kerr geometry there is no Killing vector which is 
timelike everywhere in the region exterior of the outer horizon. 
In his proof, he skillfully used differential and integral transformations 
of solutions of perturbation equations, and obtained a conserved quantity 
which is well-defined for unstable modes which are purely incoming on 
the outer horizon and purely outgoing at the infinity, 
and are characterized by positive imaginary part of their frequencies.
Then he showed that the positivity of the quantity bounds the magnitudes 
of the time derivative of perturbations. 

On the other hand, it was shown that the perturbation equations 
in the Kerr geometry are obtained from those in the Kerr-de Sitter geometry 
in the confluent limit.
In other words, a irregular singularity at infinity of the equation 
in the Kerr case is separated into two regular singularities 
of the equation in the Kerr-de Sitter geometry by cosmological constant.
In a series of paper~\cite{STU1,STU2,STU3}, 
Suzuki, Takasugi and the author have constructed analytic 
solutions of the perturbation equations of massless fields in 
the Kerr-de Sitter geometries.
We found transformations such that both the angular and the radial equations 
are reduced to Heun's equation~\cite{STU1}.
The solution of Heun's equation (Heun's function) is expressed in the form 
of a series of hypergeometric functions, and its coefficients are determined 
by three-term recurrence relations~\cite{BookHeun}.
The solution of the radial equation which is valid in the entire physical 
region is obtained by matching two solutions which have different convergence 
regions in the region where both solutions are convergent.
We examined properties of the solution in detail and, in particular, 
analytically showed~\cite{STU2} that our solution satisfies 
the Teukolsky-Starobinsky identities~\cite{T-S}.
There are similarities between the procedures for solving 
the perturbation equations in the Kerr and Kerr-de Sitter geometries.
Thus differential and integral transformations of Heun's function 
may be useful for studying mode stability of the Kerr-de Sitter geometry.

In this paper, we investigate differential and integral transformations 
of solutions of massless perturbation equations in the Kerr-de Sitter geometry.
These transformations map a solution of a Heun's equation to a solution 
of another Heun's equation which has different parameters from original ones.
Differential transformations include the Teukolsky-Starobinsky identities 
as special cases, and also other transformations for angular functions.
We will consider two specific integral transformations because we do not 
know systematic way to study integral transformations of Heun's function. 
Then we will apply the integral transformation to a solution of the radial 
equation. 
It will turn out to be possible to make the transformations 
of radial functions only for unstable modes which are purely incoming 
on the outer horizon and purely outgoing on the de Sitter horizon, 
and have positive imaginary part of their frequencies.
A conserved energy integral will be constructed from the transforms 
of the angular and radial functions for unstable modes, 
and we discuss properties of the quantity.
In the Kerr limit, these transformations and the conserved quantity are 
coincide with those given in Ref.~\cite{Whiting} by Whiting in order to 
prove mode stability of the Kerr black hole.

\section{The Teukolsky equations for the Kerr-de Sitter \\ geometry}
\hspace*{\parindent}
\setcounter{equation}{0}
We consider perturbation equations for massless fields in 
the Kerr-de Sitter geometries.
In the Boyer-Lindquist coordinates, the Kerr-de Sitter metric has the form,
\begin{eqnarray}
ds^2 &=& -\rho^2
\left(\frac{dr^2}{\Delta_r}+\frac{d\theta^2}{\Delta_\theta}\right)
-\frac{\Delta_\theta \sin^2\theta}{(1+\alpha)^2 \rho^2}
[adt-(r^2+a^2)d\varphi]^2 \nonumber \\ 
&&  +\frac{\Delta_r}{(1+\alpha)^2 \rho^2}(dt-a\sin^2\theta
d\varphi)^2,
\end{eqnarray}
where $\alpha=\Lambda a^2/3$,  
$\rho^2=\bar{\rho}\bar{\rho}^*$, $\bar{\rho}=r+ia\cos\theta$ and 
\begin{eqnarray}
\Delta_r&=&(r^2+a^2)\left(1-\frac{\alpha}{a^2}r^2\right)-2Mr
\nonumber\\
&=&-\frac{\alpha}{a^2}(r-r_+)(r-r_-)(r-r'_+)(r-r'_-)\;,
\nonumber\\
\Delta_\theta&=&1+\alpha\cos^2\theta\;.
\end{eqnarray}
Here $\Lambda$ is the cosmological constant, $M$ is 
the mass of the black hole, and $aM$ is its angular momentum.

We assume that the coordinate dependences of the perturbation of the field 
have form $\Phi_s=e^{-i(\omega t-m\varphi)}R_s(r)S_s(\theta)$.
Then angular Teukolsky equation for massless field with spin weight $s$ 
is given by 
\begin{eqnarray}
\Bigg\{&& \makebox[-0.7cm]{} \frac{d}{dx} ( 1+\alpha x^2 ) (1-x^2) 
   \frac{d}{dx}+ \lambda_s  
       + \frac{(1+\alpha)^2}{\alpha} \xi^2
         - 2 \alpha x^2 \nonumber \\  
 &+& \frac{1+\alpha}{1+\alpha x^2} 
     \left[ \ 2s (\alpha m - (1+\alpha) \xi) x 
          - \frac{(1+\alpha)^2}{\alpha} \xi^2 
            +2 m (1+\alpha) \xi + s^2 \ \right] \nonumber \\ 
 &-& \frac{(1+\alpha)^2 m^2}{(1+\alpha x^2) (1-x^2)}
    - \frac{(1+\alpha) (s^2 + 2smx)}{1-x^2} \ \ \Bigg\} S_s(x) = 0, 
  \label{eqn:sx}
\end{eqnarray}
where $x = \cos \theta$ and $\xi = a \omega$.
The separation constant $\lambda_s$ is an even function of $s$ 
as shown in Ref.\cite{STU1}.
This equation has five regular singularities at $\pm 1$, 
$\pm \frac{i}{\sqrt{\alpha}}$ and $\infty$.
We define the variable $z$ by 
\begin{equation}
 \label{eqn:x-z}
z = \frac{1-\frac{i}{\sqrt{\alpha}}}{2} 
    \frac{x+1}{x-\frac{i}{\sqrt{\alpha}}}.
\end{equation}
Then the singularities are transformed to $z=0, 1, z_s, z_\infty$, 
and $\infty$ where 
$z_s=-\frac{i(1+i\sqrt{\alpha})^2}{4\sqrt{\alpha}}$ 
and $z_\infty=-\frac{i(1+i\sqrt{\alpha})}{2\sqrt{\alpha}}$.
The singularity at $z=z_\infty$ which corresponds to $x=\infty$
can be factored out by the transformation 
\begin{equation}
S_s(z)=z^{C_1} (z-1)^{C_2} (z-z_s)^{C_3} (z-z_\infty) f_S(z),\label{eqn:Sf}
\end{equation}
where $C_1=\delta_1 (m-s)/2$, 
$C_2=\delta_2 (m+s)/2$ and 
$C_3=\delta_3 \frac{i}{2}\left(\frac{1+\alpha}{\sqrt{\alpha}}\xi
-\sqrt{\alpha}m-is\right)$, ($\delta_1, \delta_2, \delta_3=\pm 1$).
Now $f_S(z)$ satisfies the equation
\begin{equation}
\left\{ \ \frac{d^2}{dz^2} + \left[\frac{2C_1+1}{z} + \frac{2C_2+1}{z-1}
  +\frac{2C_3+1}{z-z_s} \right] \frac{d}{dz}
    + \frac{\rho_+ \rho_- z + u}{z (z-1)(z-z_s)} \ \right\} f_S(z) =0,
\label{eqn:HS}
\end{equation}
where 
\begin{eqnarray}
\rho_{\pm}\makebox[-3mm]{}&=&\makebox[-3mm]{}
          C_1+C_2+C_3 \pm C_3^* +1, \\
u\makebox[-3mm]{}&=&\makebox[-3mm]{}
       \frac{-i}{4\sqrt{\alpha}} \Bigg\{
            \lambda_s-2i\sqrt{\alpha}+2(1+\alpha)(m+s)\xi
            -(1+i\sqrt{\alpha})^2 (2C_1 C_2+C_1+C_2) \nonumber \\
&&\makebox[1cm]{}
     - 4i\sqrt{\alpha}(2C_1 C_3+C_1+C_3)
         -\frac{m^2}{2}\left[4\alpha+(1+i\sqrt{\alpha})^2\right] \nonumber \\
&&\makebox[1cm]{} +\frac{s^2}{2}(1-i\sqrt{\alpha})^2
      +2ims\sqrt{\alpha}(1+i\sqrt{\alpha}) \Bigg\}.
\end{eqnarray}
Equation (\ref{eqn:HS}) is called the Heun's equation 
which has four regular singularities.

Next we consider the radial Teukolsky equation which is given by 
\begin{eqnarray}
&&\makebox[-10mm]{}\Bigg\{ \ 
  \Delta_r^{-s}\frac{d}{dr}\Delta_r^{s+1}\frac{d}{dr}
    +\frac{1}{\Delta_r}\left[ (1+\alpha)^2 K^2
        - is(1+\alpha) K \frac{d\Delta_r}
            {dr} \right] \nonumber \\
&& \makebox[-5mm]{}
   +4is(1+\alpha)\omega r -\frac{2\alpha}{a^2}(s+1)(2s+1) r^2
      +s(1-\alpha) -\lambda_s \ \Bigg\} R_s(r) = 0, \label{eqn:Rr}
\end{eqnarray}
with $K=\omega(r^2+a^2)-am$.
This equation has five regular singularities at 
$r_\pm, r'_\pm$, and $\infty$ which are assigned such that  
$r_\pm \rightarrow M\pm\sqrt{M^2-a^2-Q^2} \equiv r^0_\pm$ and 
$r'_\pm \rightarrow \pm a/\sqrt{\alpha}$ 
in the limit $\alpha \rightarrow 0 (\Lambda \rightarrow 0)$. 
And the coefficients of the equation  are complex for spin fields ($s\neq 0$).
We assume that the cosmological constant is sufficiently small 
so that all $r_\pm, r'_\pm$ are real.
By using the new variable 
\begin{equation}
 \label{eqn:z-r}
z=\left (\frac{r_+ - r'_-}{r_+ - r_-}\right )
\left ( \frac{r - r_-}{r - r'_-}\right ),
\end{equation}
Eq.(\ref{eqn:Rr}) becomes an equation which has regular singularities at 
$0, 1, z_r, z_\infty$ and $\infty$,
$$
z_r=\left(\frac{r_+ - r'_-}{r_+ - r_-}\right )\left(
     \frac{r'_+ - r_-}{r'_+ - r'_-}\right)\;,
\quad\quad
z_\infty=\frac{r_+ - r'_-}{r_+ - r_-}.
$$
To proceed further, we define the parameters 
\begin{eqnarray}
D_{i\pm}&=&\frac{1}{2}\left\{-s \pm (2a_i+s  )\right\}. \qquad 
    (i=1,2,3,4)
\end{eqnarray}
Here $a$'s are purely imaginary numbers defined by 
\begin{eqnarray}
a_1&=&i\frac{a^2(1+\alpha)\left(\omega(r_+^2 + a^2)-am \right)}
            { \alpha (r'_+ - r_+)(r'_- - r_+)(r_- - r_+)},\nonumber \\
a_2&=&i\frac{a^2(1+\alpha)\left(\omega(r_-^2 + a^2)-am \right)}
            { \alpha (r'_+ - r_-)(r'_- - r_-)(r_+ - r_-)}, \nonumber \\
a_3&=&i\frac{a^2(1+\alpha)\left(\omega({r'_+}^2 + a^2)-am \right)}
            { \alpha (r_- - r'_+)(r'_- - r'_+)(r_+ - r'_+)},\nonumber \\
a_4&=&i\frac{a^2(1+\alpha)\left(\omega(r_-'^2 + a^2)-am \right)}
            { \alpha (r_- - r'_-)(r'_+ - r'_-)(r_+ - r'_-)}, 
\end{eqnarray}
and the relation $a_1+a_2+a_3+a_4=0$ is satisfied.
Again we can factor out the singularity at $z=z_\infty$ by the 
transformation as 
\begin{equation}
R_s(z)=z^{D_2} (z-1)^{D_1} (z-z_r)^{D_3} (z-z_\infty)^{2s+1} f_R(z),
\end{equation}
where $D_i \ (i=1,2,3)$ is either $D_{i+}$ or $D_{i-}$. 
Then, $f_R(z)$ satisfies the Heun's  equation as
\begin{equation}
\left\{ \ \frac{d^2}{dz^2} + \left[\frac{\gamma}{z} + \frac{\delta}{z-1}
   +\frac{\epsilon}{z-z_r} \right] \frac{d}{dz} 
   + \frac{\sigma_+ \sigma_- z + v}{z (z-1)(z-z_r)} \ \right\} f_R(z) =0,
\label{eqn:HR}
\end{equation}
where 
\begin{eqnarray}
\gamma=2D_2+s+1, \quad && \delta=2D_1+s+1, 
            \qquad \epsilon=2D_3+s+1, \nonumber \\
\sigma_\pm &=& 
  D_1 + D_2 + D_3 +D_{4\mp}+ 2s+1.
\end{eqnarray}
The parameters $\gamma$, $\delta$, $\epsilon$ and $\sigma_{\pm}$ 
satisfy the following relation, 
\begin{equation}
\gamma+\delta-1=\sigma_+ +\sigma_- -\epsilon,
\end{equation}
which is required for Eq.(\ref{eqn:HR}) to be a Heun's equation.
The remaining parameter $v$ is given by
\begin{eqnarray}
v &=&   \frac{2a^4 (1+\alpha)^2}{\alpha^2 {\cal D}}
        \frac{(r_+ - r'_+)^2 (r_+ - r'_-)^2 (r_- - r'_-) (r'_+ - r'_-)}
             {r_+ - r_-} \nonumber \\
&& \makebox[-8mm]{}
   \times \Bigg\{-\omega^2 r_-^3(r_+ r_- - 2r_+ r'_+ + r_- r'_+)
     +2a\omega(a\omega-m)r_- (r_+ r'_+ - r_-^2) \nonumber \\
&& -a^2(a\omega-m)^2(2r_- - r_+ - r'_+) \nonumber \\
&& +\frac{2isa^2(1+\alpha)}{\alpha}
     \frac{\left[ \omega (r_- r'_- +a^2)-am \right]}
          {(r_+ - r_-)(r'_+ - r'_-)(r_- - r'_-)} \nonumber \\
&& +(s+1)(2s+1)\left[\frac{2{r'_-}^2}
        {(r_+ - r_-)(r'_+ -r'_-)}-z_\infty\right]
\nonumber \\
&& - 2D_2(z_r D_1+D_3)-(s+1)\left[ (1+z_r)D_2+z_r D_1+D_3 \right] \nonumber \\
&& -\frac{a^2}{\alpha(r_+ - r_-)(r'_+ - r'_-)} 
\left[-\lambda_s+s(1-\alpha)\right] \Bigg\}.
\end{eqnarray}
Here ${\cal D}$ is the discriminant of $\Delta_r = 0$,
\begin{eqnarray*}
{\cal D}&=&(r_+ - r_-)^2 (r_+ - r'_+)^2 (r_+ - r'_-)^2
            (r_- - r'_+)^2 (r_- - r'_-)^2 (r'_+ - r'_-)^2 \\
&=& \frac{16a^{10}}{\alpha^5}\Bigg\{ \ 
        (1-\alpha)^3\left[M^2-(1-\alpha)a^2\right]  \\
&&  +\frac{\alpha}{a^2}\left[-27M^4+36(1-\alpha)M^2a^2
       - 8(1-\alpha)^2a^4\right]
           -\frac{16\alpha^2}{a^4}a^6 \ \Bigg\}.
\end{eqnarray*}

It should be noted that the equation (\ref{eqn:Rr}) for $s=0$ is 
the perturbation equation for a conformal scalar field 
which satisfies $\Box \phi=\frac16R \phi$.
In the case of an ordinary scalar field which satisfies $\Box \phi=0$, 
the term $-\frac{2\alpha}{a^2} r^2$ in Eq.(\ref{eqn:Rr}) is absent.


\section{Differential transformations}
\hspace*{\parindent}
\setcounter{equation}{0}
We consider a differential transformation of a Heun's function $f(z)$ 
which satisfies 
\begin{eqnarray}
 \label{eqn:HE}
&& \hspace{-10mm}
   M_z(\gamma, \delta, \epsilon; \alpha, \beta; q) \ f(z) \equiv
    \Bigg\{ \ z(z-1)(z-a_H)\frac{\partial^2}{\partial z^2} \nonumber \\ 
&& \hspace{3mm}   + \left[\gamma (z-1)(z-a_H) + \delta z(z-a_H)
    +\epsilon z(z-1)\right] \frac{\partial}{\partial z}
    +\alpha\beta z + q \ \Bigg\} f(z) =0.
\end{eqnarray}
Differentiating this equation $N$ times, we obtain 
\begin{eqnarray}
&& \hspace{-5mm}
 \left(\frac{d}{dz}\right)^N 
  M_z(\gamma, \delta, \epsilon; \alpha, \beta; q) \ f(z)=
   \Bigg\{ z(z-1)(z-a_H)\frac{\partial^2}{\partial z^2} \nonumber \\
&& +\left[(\gamma+N)(z-1)(z-a_H)+(\delta+N)z(z-a_H)
     +(\epsilon+N)z(z-1)\right]\frac{\partial}{\partial z} \nonumber \\
&& +\left[3N^2+(2\alpha+2\beta-1)N+\alpha\beta\right]z+
    q-N(N-1)(a_H+1)-N((a_H+1)\gamma+a_H\delta+\epsilon) \nonumber \\
&& +N(N-1+\alpha)(N-1+\beta)\left(\frac{\partial}{\partial z}\right)^{-1} 
    \Bigg\}\left(\frac{d}{dz}\right)^N f(z).
\end{eqnarray}
Therefore 
\begin{equation}
\tilde{f}(z)=\left(\frac{d}{dz}\right)^N f(z)
\end{equation}
with $N=1-\alpha$ or $1-\beta$ formally satisfies 
another Heun's equation in which the parameters $(\gamma, \delta, \cdots)$ 
are replaced by 
$\tilde{\gamma}=\gamma+N, \tilde{\delta}=\delta+N, 
\tilde{\epsilon}=\epsilon+N, 
\tilde{q}=q-N(N-1)(a_H+1)-N((a_H+1)\gamma+a_H\delta+\epsilon)$, 
and $\tilde{\alpha}, \tilde{\beta}$ 
which are determined by relations 
$\tilde{\alpha}+\tilde{\beta} =\alpha+\beta+3N$ and 
$\tilde{\alpha}\tilde{\beta}=3N^2+(2\alpha+2\beta-1)N+\alpha\beta$. 
Here we choose $N=1-\alpha$.
Properly $N$ should be an positive integer.
In both angular and radial cases, we can take parameters so that $N=|2s|$.
These transformations are the Teukolsky-Starobinsky relations~\cite{T-S}.
In Ref.\cite{STU2} it was shown that the radial Teukolsky-Starobinsky 
relations are satisfied by using an analytic solution expressed by the form 
of a series of hypergeometric functions.

In the angular case, $N=|s-m|$ and $N=|s+m|$ are also possible.
We now explain the case with $\delta_1=\delta_2=-\delta_3$ 
in the definitions of $C_i \ (i=1,2,3)$, explicitly. 
Then $N=\delta_1(s-m)$ and thus we can take $\delta_1(s-m)=|s-m|$.
The transform of angular function 
$\tilde{f}_S(z)=\left(\frac{d}{dz}\right)^N f_S(z)$ satisfies 
the Heun's equation with the following parameters, 
\begin{eqnarray}
&& \tilde{\gamma}=1, \qquad \tilde{\delta}=2\delta_1 s+1, \qquad 
   \tilde{\epsilon}=-\delta_1\left[i\frac{1+\alpha}{\sqrt{\alpha}}\xi
                   +(1-i\sqrt{\alpha})m\right]+1, \nonumber \\
&& \tilde{\rho}_+ = 1+|s-m|, \qquad 
   \tilde{\rho}_- = -\delta_1 i \left[\frac{1+\alpha}{\sqrt{\alpha}}\xi
                     -\sqrt{\alpha}m+is\right]+1, \\
&& \tilde{u}=u-|s-m|\left[z_s(\delta_1(m+s)+1)
               -\delta_1 i \left(\frac{1+\alpha}{\sqrt{\alpha}}\xi
                  -\sqrt{\alpha}m -is \right)+1\right]. \nonumber 
\end{eqnarray}
We next define $\tilde{S}_s(z)$ from $\tilde{f}_S(z)$ through the inverse 
procedures to those used to derive $f_S(z)$ from $S_s(z)$.
By setting $\tilde{\gamma}=2\tilde{C}_1+1, \tilde{\delta}=2\tilde{C}_2+1$ and 
$\tilde{\epsilon}=2\tilde{C}_3+1$, we obtain 
$\tilde{C}_1=0, \tilde{C}_2=\delta_1 s$ and 
$\tilde{C}_3=-\frac{\delta_1}{2}\left[i\frac{1+\alpha}{\sqrt{\alpha}}\xi
                    +(1-i\sqrt{\alpha})m \right]$.
Using these $\tilde{C}_i$ as the exponents at the singularities, 
$\tilde{S}_s(z)$ is given by  
\begin{equation}
\tilde{S}_s(z)\equiv z^{\tilde{C}_1}(z-1)^{\tilde{C}_2}(z-z_s)^{\tilde{C}_3}
                   (z-z_\infty)\tilde{f}_S(z).
\end{equation}
This new angular function $\tilde{S}_s$ satisfies a similar 
equation to the angular Teukolsky equation (\ref{eqn:sx}) 
satisfied by $S_s$;
\begin{eqnarray}
 \label{eqn:transf-ang}
&& \Bigg\{ \frac{d}{dx}(1-x^2)(1+\alpha x^2)\frac{d}{dx}
         +\lambda_s-2\alpha x^2 
         -(1+\alpha)^2 \frac{1-x^2}{1+\alpha x^2}\xi^2
         +2(1+\alpha)^2\frac{1-x}{1+\alpha x^2}m\xi \nonumber \\
&& \hspace{10mm}
          -\alpha(1+\alpha)\frac{(1-x)^2}{1+\alpha x^2}m^2
         -(1+\alpha)\frac{1+x}{1-x}s^2 \Bigg\}\tilde{S}_s(x)=0,
\end{eqnarray} 
where we used (\ref{eqn:x-z}).
We note that the operator acting on $\tilde{S}_s$ 
in the above equation is invariant under $s \longrightarrow -s$ 
because $\lambda_s$ is an even function of $s$.


\section{Integral transformations}
\hspace*{\parindent}
\setcounter{equation}{0}
We construct integral transformations of Heun's function $f(z)$ 
which satisfies Eq.(\ref{eqn:HE}).
The integral transformation which maps a solution of a Heun's equation 
to a solution of another Heun's equation is given by 
\begin{equation}
 \label{eqn:int1}
\tilde{f}(z)=\int_{{\cal C}} dt \ t^{\gamma-1}(t-1)^{\delta-1}
                (t-a_H)^{\epsilon-1}K(z, t)f(t).  \qquad({\rm type \ A})
\end{equation}
The function $\tilde{f}(z)$ is a solution of 
$M_z (\tilde{\gamma}, \tilde{\delta}, \tilde{\epsilon}; 
       \tilde{\alpha}, \tilde{\beta}; \tilde{q}) \ \tilde{f}(z)=0$
if the kernel satisfy the condition 
\begin{equation}
 \label{eqn:Mz-Mt}
\left[
  M_z (\tilde{\gamma}, \tilde{\delta}, \tilde{\epsilon}; 
     \tilde{\alpha}, \tilde{\beta}; \tilde{q})
 -M_t (\gamma, \delta, \epsilon; \alpha, \beta; q)
  \right] K(z, t)=0,
\end{equation}
and the surface term of the integral 
\begin{equation}
 \label{eqn:Wron}
t^{\gamma}(t-1)^{\delta}(t-a_H)^{\epsilon}
  \left\{ K(z, t)\frac{\partial f(t)}{\partial t}
        -\frac{\partial K(z, t)}{\partial t}f(t)\right\}
   \equiv W(z, t),
\end{equation}
vanishes at the ends of ${\cal C}$.

In this paper we consider the kernels which depend on $z$ 
and $t$ only by the form of their product $zt\equiv \zeta$ 
thus $K(z, t)=K(\zeta)$.
Then Eq.(\ref{eqn:Mz-Mt}) can be rewritten as 
\begin{eqnarray}
&& \hspace{-1mm}
   \Bigg\{ \ t\left[ \zeta(\zeta-a_H)\frac{d^2}{d\zeta^2}
         +\Big((\alpha+\beta+1)\zeta-a_H\tilde{\gamma}\Big)\frac{d}{d\zeta}
          +\alpha\beta \right] \nonumber \\
&& -z\left[ \zeta(\zeta-a_H)\frac{d^2}{d\zeta^2}
   +\Big((\tilde{\alpha}+\tilde{\beta}+1)\zeta-a_H\gamma\Big)\frac{d}{d\zeta}
     +\tilde{\alpha}\tilde{\beta} \right] \nonumber \\ 
&& +\Big[(1+a_H)(\tilde{\gamma}-\gamma)+a_H(\tilde{\delta}-\delta)
               +\tilde{\epsilon}-\epsilon\Big]\zeta\frac{d}{d\zeta}
    +q-\tilde{q} \ \Bigg\} K(\zeta)=0.
\end{eqnarray}
In particular, we require the kernel to satisfy the following equations,
\begin{eqnarray}
&&\left[ \zeta(\zeta-a_H)\frac{d^2}{d\zeta^2}
         +\Big((\alpha+\beta+1)\zeta-a_H\tilde{\gamma}\Big)\frac{d}{d\zeta}
          +\alpha\beta \right]K(\zeta)=0, \nonumber \\
&&\left[ \zeta(\zeta-a_H)\frac{d^2}{d\zeta^2}
   +\Big((\tilde{\alpha}+\tilde{\beta}+1)\zeta-a_H\gamma\Big)\frac{d}{d\zeta}
     +\tilde{\alpha}\tilde{\beta} \right]K(\zeta)=0, \\
&&(1+a_H)(\tilde{\gamma}-\gamma)+a_H(\tilde{\delta}-\delta)
               +\tilde{\epsilon}-\epsilon=0, \nonumber \\
&& q=\tilde{q}. \nonumber 
\end{eqnarray}
There are four sets of solutions of these equations.
Using invariance of the equations under 
$\alpha \longleftrightarrow \beta$ 
and $\tilde{\alpha} \longleftrightarrow \tilde{\beta}$, 
we can take the solutions without loss of generality as 
\begin{eqnarray}
&& \hspace{30mm} \label{eqn:kernel}
   K(\zeta)=\left(\zeta-a_H\right)^{-\beta}, \\
&& \label{eqn:tilde-para}
  \tilde{\alpha}=\gamma, \quad \tilde{\beta}=\beta, 
    \quad \tilde{\gamma}=\alpha, \quad 
      \tilde{\delta}=\beta-\epsilon+1, \quad
       \tilde{\epsilon}=\beta-\delta+1, \quad \tilde{q}=q.
\end{eqnarray}

Next we define another type of integral transformation,
\begin{equation}
 \label{eqn:int2}
\tilde{f}(z)=\int_1^{\frac{a_H}{z}} dt \ t^{\gamma-1}(t-1)^{\delta-1}
                (t-a_H)^{\epsilon-1}K(z, t)f(t), \qquad({\rm type \ B})
\end{equation}
where it should be noted that the integral region depends on $z$.
In order for $\tilde{f}(z)$ to satisfy Heun's equation, 
the following condition must hold instead of the conditions (\ref{eqn:Wron});
\begin{eqnarray}
 \label{eqn:condition}
&& \hspace{-5mm}
   W(z, t)\Big|_1^{\frac{a_H}{z}}+z(z-1)(z-a_H)\Bigg\{\frac{\partial}{\partial z}
        \bigg[\frac{a_H}{z^2}K\left(z, t\right) 
           t^{\gamma-1}(t-1)^{\delta-1}(t-a_H)^{\epsilon-1} 
             f(t)\Big|_{t=\frac{a_H}{z}} \bigg] \nonumber \\
&& +\frac{a_H}{z^2}\frac{\partial K(z, t)}{\partial z}
      t^{\gamma-1}(t-1)^{\delta-1}(t-a_H)^{\epsilon-1} f(t)
        \Bigg|_{t=\frac{a_H}{z}}\Bigg\} \\
&& \hspace{-5mm}
   +[\tilde{\gamma}(z-1)(z-a_H)+\tilde{\delta}z(z-a_H)+\tilde{\epsilon}z(z-1)] 
         \frac{a_H}{z^2} K(z, t)
           t^{\gamma-1}(t-1)^{\delta-1}(t-a_H)^{\epsilon-1}
             f(t)\Bigg|_{t=\frac{a_H}{z}} \nonumber \\
&& =0. \nonumber  
\end{eqnarray}
We can show by using the kernel in Eq.(\ref{eqn:kernel}) that 
$\tilde{f}(z)$ is a solution of the Heun's equation with the parameters 
in Eq.(\ref{eqn:tilde-para}), provided that this condition is satisfied.


\section{Conserved energy integral for unstable modes}
\hspace*{\parindent}
\setcounter{equation}{0}
In this section we are going to construct a conserved quantity 
for unstable modes in the Kerr-de Sitter geometry by using 
the differential and integral transformations of Heun's function given 
in section 3 and 4.
We will assume that the cosmological constant is sufficiently small so that 
$r'_- \ll r_- < r_+ \ll r'_+$.

First, it can be easily checked that the differential transformations of 
the angular function for the Kerr-de Sitter black hole which we provided 
in section 3 coincide with those given by Whiting in the Kerr limit 
$\Lambda \longrightarrow 0 \ (\alpha\longrightarrow 0)$.
Indeed since the Teukolsky equations for the Kerr black hole have the forms 
of confluent Heun's equation~\cite{STU1}, this differential transformation 
in the Kerr limit becomes one for the confluent Heun's function.

Next we consider the integral transformations of the radial function $f_R(z)$ 
which satisfies Eq.(\ref{eqn:HR}).
We choose $D_1=-a_1-s, D_2=-a_2-s$ and $D_3=-a_3-s$ as the parameters 
included in the radial equation (\ref{eqn:HR}).
Then we have 
\begin{eqnarray}
&&\sigma_+ = -2s+1, \qquad \sigma_- = 2a_4-s+1, \nonumber \\
&&\gamma=-2a_2-s+1, \quad \delta=-2a_1-s+1, \quad \epsilon=-2a_3-s+1.
\end{eqnarray}
The unstable modes are purely incoming on the outer horizon $z=1$ and 
purely outgoing on the de Sitter horizon $z=z_r$, 
\begin{eqnarray}
R_s(z) &\sim&  (z-1)^{-s-a_1}, \qquad (z \sim 1) \nonumber \\
       &\sim&  \left(1-\frac{z}{z_r}\right)^{a_3},  \qquad (z \sim z_r)
\end{eqnarray}
or equivalently
\begin{eqnarray}
f_R(z) &\sim&  (z-1)^0, \qquad (z \sim 1) \nonumber \\
       &\sim&  \left(1-\frac{z}{z_r}\right)^{2a_3+s},  \qquad (z \sim z_r)
\end{eqnarray}
where we used the parameters determined above.
Furthermore the unstable modes are characterized by having positive 
imaginary part of frequency $\omega$.
In our discussions below, 
we choose the integral region as ${\cal C}=(1, z_r)$ in the integral 
transformation of type A.

We first examine the Kerr limit.
In the Kerr limit, we find 
\begin{equation}
z_r \longrightarrow \frac{a}{2\sqrt{\alpha}(r_+^0 -r_-^0)}, \qquad 
a_4 \longrightarrow \frac{ia\omega}{2\sqrt{\alpha}},
\end{equation}
where $r_\pm^0=M\pm\sqrt{M^2-a^2}$.
Then both integral regions of the transformations of type A and B become 
$(1, \infty)$ and the kernel (\ref{eqn:kernel}) becomes of the Laplace type, 
\begin{equation}
K(\zeta) \sim e^{2i\omega (r_+^0 -r_-^0)\zeta}.
\end{equation}
Thus both integral transformations coincide with those used in the proof of 
mode stability of the Kerr black hole in this limit. 

We next consider the boundary term $W(z, t)\Big|_1^{z_r}$.
In type A case, $W(z, t)$ behaves as 
\begin{eqnarray}
W(z, t) &\sim& (t-1)^{-2a_1-s+1}, \hspace{35mm} (t \sim 1) \nonumber \\
        &\sim& (t-z_r)^{-2a_3-s+1}\frac{d}{dt}(t-z_r)^{2a_3+s}. 
           \qquad (t \sim z_r)
\end{eqnarray}
The real part of $a_1$ is negative if the imaginary part of $\omega$ 
is positive. 
Thus if $s\leq 0$, $W(z, t)$ vanishes at $t=1$ for unstable modes.
On the other hand, $W(z, t)$ does not vanish at $t=z_r$.
Therefore this transformation is not appropriate.
Although there is also a possibility that our choice of the integral region 
${\cal C}$ is wrong, we do not adopt the transformation of type A 
in our discussions below.
In type B case, $W(z, t)$ behaves as 
\begin{eqnarray}
W(z, t) &\sim& (t-1)^{-2a_1-s+1}, \hspace{13mm} (t \sim 1) \nonumber \\
        &\sim& \left(t-\frac{z_r}{z}\right)^{-2a_4+s-1}. 
            \qquad \left(t \sim \frac{z_r}{z}\right)
\end{eqnarray}
Since the real part of $a_4$ is negative for unstable modes, 
it is clear that the condition (\ref{eqn:condition}) holds 
for unstable modes from the form of the kernel $K(z, t)=(zt-z_r)^{-2a_4+s-1}$.
Hence it is possible to make the integral transformation of type B 
for radial functions of unstable modes.

The radial function $\tilde{f}_R(z)$ given by the integral transformation of 
type B of $f_R(z)$, 
\begin{equation}
\tilde{f}_R(z)=\int_1^{\frac{z_r}{z}} dt \ t^{-2a_2-s}(t-1)^{-2a_1-s}
                (t-z_r)^{-2a_3-s}(zt-z_r)^{-2a_4+s-1}f_R(t),
\end{equation}
satisfies the Heun's equation with parameters (\ref{eqn:tilde-para}), 
\begin{eqnarray}
&& \tilde{\sigma}_+ =-2a_2-s+1, \hspace{17mm}
      \tilde{\sigma}_- =2a_4-s+1, \nonumber \\
&& \tilde{\gamma}=-2s+1\equiv 2\tilde{D}_2+1, \qquad 
   \tilde{\delta}=-2a_1-2a_2+1\equiv 2\tilde{D}_1+1, \nonumber \\ 
&& \tilde{\epsilon}=-2a_2-2a_3+1\equiv 2\tilde{D}_3+1, \qquad 
   \tilde{v}=v.
\end{eqnarray}
We define a new radial function $\tilde{R}_s(z)$ from $\tilde{f}_R$ by 
\begin{equation}
\tilde{R}_s(z)=z^{\tilde{D}_2}(z-1)^{\tilde{D}_1}
               (z-z_r)^{\tilde{D}_3}(z-z_\infty)\tilde{f}_R(z).
\end{equation}
This function satisfies the following equation 
\begin{equation}
\left\{\frac{d}{dr}\Delta_r\frac{d}{dr}-\frac{2\alpha}{a^2}
     -s^2 F_{s}(r) -m^2 F_m(r) +\omega^2 F_\omega(r)
      +m\omega F_{m\omega}(r) -\lambda_s \right\}\tilde{R}_s(r)=0,
\end{equation}
where 
\begin{eqnarray}
F_s(r) &=& -\frac{\alpha}{a^2}(r'_+ -r_-)(r'_- -r_-)\frac{r-r_+}{r-r_-}
            +\frac{\alpha}{a^2}(r_+ +r_-)^2, \nonumber \\
F_m(r) &=& \frac{4a^4(1+\alpha)^2}
                {\alpha^2(r_+ -r_-)^2(r'_+ -r_-)^2(r'_- -r_-)^2} \cdot 
           \frac{(r-r_-)^2}
                {(r-r_+)(r-r'_+)(r-r'_-)} \nonumber \\
&&   \times \bigg\{\alpha(r_+ -r_-)(r_+ +r_-)^2
       +\left[(1-\alpha)a^2-2\alpha r_-^2\right](r-r_+)\bigg\}, \nonumber \\
F_\omega(r) &=& 
    -\frac{a^2(1+\alpha)^2}
          {\alpha^3(r_+ -r_-)^2(r'_+ -r_-)^2(r'_- -r_-)^2} \cdot
     \frac{(r-r_-)}{(r-r_+)(r-r'_+)(r-r'_-)} \nonumber \\
&& \hspace{-15mm}\times 
   \bigg\{(r_+^2 -r_-^2)^2\left[a^2(1+\alpha)
     -\alpha(r_+ +r_-)^2\right]^2 \nonumber \\
&& \hspace{-12mm}
    +2(r_+ -r_-)\Big[2(1-\alpha)a^6
     +a^4\Big((1-4\alpha+5\alpha^2)r_+^2-4\alpha(1-2\alpha)r_+ r_- 
       +(1-8\alpha+5\alpha^2)r_-^2\Big)\nonumber \\
&& \hspace{-8mm} 
   -2\alpha a^2\left((1-\alpha)r_+^4 +2(1-\alpha)r_+^3 r_- 
      +(3-7\alpha)r_+^2 r_-^2 -8\alpha r_+ r_-^3 
       +(1-5\alpha)r_-^4\right) \nonumber \\ 
&& \hspace{-8mm}
    +\alpha^2(r_+^6 +4r_+^5 r_- +9r_+^4 r_-^2 +8r_+^3 r_-^3 
       +5r_+^2 r_-^4 +r_-^6)\Big](r-r_+) \nonumber \\
&&   \hspace{-12mm}
   +\Big[4\alpha(1-\alpha)a^6+a^4\Big((1-\alpha)^2r_+^2-2(1-\alpha)^2r_+ r_-
       +(1+6\alpha-15\alpha^2)r_-^2\Big) \nonumber \\ 
&& \hspace{-8mm} 
   -2\alpha a^2\Big((1-\alpha)r_+^4-4(1-\alpha)r_+ r_-^3
      +(1+7\alpha)r_-^4 \Big) \nonumber \\
&& \hspace{-8mm} 
    +\alpha^2(r_+^6+2r_+^5 r_- +3r_+^4 r_-^2 -4r_+^3 r_-^3 
      -5r_+^2 r_-^4 -6r_+ r_-^5 +r_-^6)\Big](r-r_+)^2 \bigg\}, \nonumber \\
F_{m\omega}(r) &=& 
  -\frac{4a^3(1+\alpha)^2}
          {\alpha^3(r_+ -r_-)^2(r'_+ -r_-)^2(r'_- -r_-)^2} \cdot
     \frac{(r-r_-)}{(r-r_+)(r-r'_+)(r-r'_-)} \nonumber \\
&& \hspace{-5mm} \times \bigg\{ 
     \alpha(r_+^2-r_-^2)^2\left[-(1+\alpha)a^2+\alpha(r_+ -r_-)^2\right] 
        \nonumber \\
&&  +(r_+ -r_-)\Big[-(1-\alpha)a^4-4\alpha^2(r_+ +r_-)^2 r_-^2 
      \nonumber \\
&& \quad  +\alpha a^2\Big((1-3\alpha)(r_+ +r_-)^2 +2(1+\alpha)r_-^2\Big)
      \Big](r-r_+) \nonumber \\
&& -2\alpha(a^2+r_-^2)\left[a^2(1-\alpha)-2\alpha r_-^2\right]
     (r-r_+)^2  \bigg\},
\end{eqnarray}
Here we used Eq.(\ref{eqn:z-r}) for rewriting the equation in terms of $r$.
All the coefficients of this equation are real for any spin weight $s$ 
in contrast with the radial Teukolsky equation (\ref{eqn:Rr}).
We note that $\tilde{R}_s(r)$ behaves near $r\sim r_+$ and $r'_+$ as 
\begin{eqnarray}
 \label{eqn:behavior}
\tilde{R}_s(r) 
  &\sim& (r-r_+)^{-a_1-a_2}, \hspace{15mm} (r\sim r_+)  \nonumber \\
  &\sim& (r-r'_+)^{-a_1-a_4}. \qquad (r\sim r'_+)
\end{eqnarray}

We construct the function $\tilde{\Phi}_s$ from $\tilde{S}_s(\theta)$ and 
$\tilde{R}_s(r)$ as
\begin{equation}
\tilde{\Phi}_s=e^{-i(\omega t-m\varphi)}\tilde{R}_s(r)\tilde{S}_s(\theta),
\end{equation}
then this function satisfies 
\begin{eqnarray}
\label{eqn:transf-eqn}
&& \Bigg\{\frac{\partial}{\partial r}\Delta_r\frac{\partial}{\partial r}
       +\frac{1}{\sin\theta}\frac{\partial}{\partial\theta}
          \sin\theta(1+\alpha\cos^2\theta)
            \frac{\partial}{\partial\theta}
   -\left[F_\omega(r)
     -(1+\alpha)^2a^2\frac{\sin^2\theta}{1+\alpha\cos^2\theta}\right]
       \frac{\partial^2}{\partial t^2} \nonumber \\
&& \quad +\left[F_m(r)
   +\alpha(1+\alpha)\frac{(1-\cos\theta)^2}{1+\alpha\cos^2\theta}\right]
    \frac{\partial^2}{\partial \varphi^2} 
   +\left[F_{m\omega}
     +2(1+\alpha)^2 a \frac{1-\cos\theta}{1+\alpha\cos^2\theta}\right]
     \frac{\partial^2}{\partial t\partial\varphi} \nonumber \\
&& \quad -s^2\left[F_s(r)
      +(1+\alpha)\frac{1+\cos\theta}{1+\alpha\cos^2\theta}\right]
       -2\alpha\left(\frac{r^2}{a^2}+\cos^2\theta\right)
   \Bigg\} \tilde{\Phi}_s=0.
\end{eqnarray}
All the coefficient in this equation are real and invariant 
under $s \longrightarrow -s$.

Finally we obtain a conserved energy integral from this equation 
in the following form:
\begin{eqnarray}
&& \hspace{-10mm} 
  \int dr d\theta d\varphi \sin\theta
   \Bigg\{ \left[F_\omega(r)
     -(1+\alpha)^2a^2\frac{\sin^2\theta}{1+\alpha\cos^2\theta}\right]
       \left|\frac{\partial\tilde{\Phi}_s}{\partial t}\right|^2 
    +\Delta_r\left|\frac{\partial\tilde{\Phi}_s}{\partial r}\right|^2 
      \nonumber \\
&& \hspace{-5mm}    
   +(1+\alpha\cos^2\theta) 
      \left|\frac{\partial\tilde{\Phi}_s}{\partial\theta}\right|^2 
   +\left[F_m(r)
   +\alpha(1+\alpha)\frac{(1-\cos\theta)^2}{1+\alpha\cos^2\theta}\right]
      \left|\frac{\partial\tilde{\Phi}_s}{\partial\varphi}\right|^2
           \nonumber \\
&&  + s^2\left[F_s(r)
      +(1+\alpha)\frac{1+\cos\theta}{1-\cos\theta}\right]
       \left|\tilde{\Phi}_s\right|^2
       +2\alpha\left(\frac{r^2}{a^2}+\cos^2\theta\right)
        \left|\tilde{\Phi}_s\right|^2 \Bigg\},
\end{eqnarray}
where $r$ integration is performed over $(r_+, r'_+)$.
From Eq.(\ref{eqn:behavior}), it can be understood that this integration 
is finite for unstable modes, provided that the cosmological constant 
is sufficiently small.


\section{Summary and discussions} 
\hspace*{\parindent}
\setcounter{equation}{0}
Solutions of the perturbation equations of massless fields 
in the Kerr-de Sitter geometries can be obtained by using Heun's functions.
In this paper, we constructed the differential and integral transformations 
of Heun's function.
The differential transformations of the angular and radial functions 
include the Teukolsky-Starobinsky identities as special cases.
Although we don't know generic way to study integral transformation 
of Heun's function, we provided two types of integral transformations.
In the Kerr limit, these differential and integral transformations coincide 
with those considered in the case of perturbations 
for the Kerr geometries~\cite{Whiting}.
From the equation satisfied by the transform of the perturbation, 
we have succeeded in obtaining a conserved energy integral for unstable modes.

In the proof of mode stability of the Kerr black hole, 
it is crucial that the conserved quantity which is obtained by 
similar procedures to those used in this paper is positive definite.
From the positivity, it is concluded that the value of the conserved quantity 
bounds the integral of $\left|\frac{\partial\Phi_s}{\partial t}\right|^2$
and thus unstable modes cannot exist.
However the conserved quantity obtained for the Kerr-de Sitter black hole 
is not positive definite for the sufficiently small ( but non-vanishing ) 
cosmological constant because the coefficient of 
$\left|\frac{\partial\Phi_s}{\partial \varphi}\right|^2$ in the quantity, 
which vanishes in the Kerr limit $\Lambda \longrightarrow 0$ and 
the Schwarzschild-de Sitter limit $a \longrightarrow 0$, can become negative 
in the situation considered here.
Therefore we cannot rule out possibility that there are unstable modes.
We also point out that although $\frac{\partial}{\partial t}$ 
is globally null in the metric derived from the equation 
in the Kerr geometry~\cite{Whiting} which can be obtained 
from Eq.(\ref{eqn:transf-eqn}) in $\alpha \longrightarrow 0$ limit, 
it does not hold in the metric derived from Eq.(\ref{eqn:transf-eqn}) with 
non-vanishing cosmological constant.
However we think that the procedures performed here are natural extensions 
of those used in the Kerr case. 
It may be possible to improve the analysis which we provided here.
To the end, we think that the systematic study of integral transformations 
of Heun's function will be required. 

The analyses given in our previous papers~\cite{STU1,STU2,STU3} and here 
are applicable to the case of the Kerr-anti-de Sitter geometry similarly.
We hope that those analyses may give deeper insight for 
the correspondence between quantum gravity in anti-de Sitter space 
and conformal field theory defined on the boundary~\cite{AdS-CFT}.

\vspace{20mm}

\noindent{\Large{\bf Acknowledgments}}

I would like to thank E. Takasugi for discussions.
I would like to thank H. Suzuki for discussions and 
careful reading of this manuscript.
I would also like to thank the members of high energy theory group 
in Osaka University.


\end{document}